\documentclass[12pt]{iopart}

\usepackage{iopams}
\begin{document}

\title[Can hyperbolic phase of Brans-Dicke field account for Dark Matter?]{Can hyperbolic phase of Brans-Dicke field account for Dark Matter?}

\author{M. Ar\i k, M. \c{C}al\i k and F. \c{C}ifter}

\address{Bo\~{g}azi\c{c}i Univ., Dept. of Physics, Bebek, Istanbul, Turkey}
\address{Dogus Univ., Dept. of Sciences, Acibadem, Zeamet Street No: 21
34722 Kadikoy, Istanbul, Turkey}
\eads{\mailto{metin.arik@boun.edu.tr},
\mailto{mcalik@dogus.edu.tr},
\mailto{fcifter@dogus.edu.tr}}
\begin{abstract}
 We show that the introduction of
a hyperbolic phase for Brans-Dicke (BD) field results in a flat
vacuum cosmological solution of Hubble parameter H and fractional rate of change of BD
scalar field, F which asymptotically approach constant values. At late stages, hyperbolic phase of BD field behaves
like dark matter.

\end{abstract}

\maketitle

\section{Introduction}
It has always been one of the most challenging and interesting
problems of cosmology what the composition of the universe exactly
is: what was it in the primordial time and what is it in today's
universe? Where did the structure of the universe originally come
from? After the development of the inflationary theory
\cite{A.guth}, both observational and theoretical studies have
been continuing on this subject. According to inflationary
Universe models \cite{Robert}, inflation is capable of explaining
not only the acceleration of the expansion rate but also flatness,
homogeneity and isotropy of the universe. In addition, the
discovery of the cosmic microwave background \cite{Spergel}
indicates that our universe is nearly flat and expands with a slow
accelerating rate \cite{bernardis}-\cite{T. Roy}. This slow rate
acceleration of universe results from an adequate negative
pressure of dark energy and recent observations indicate that dark
energy behaves like Einstein's cosmological constant \cite{pierre}
which arises from the vacuum energy. The remaining energy density
is composed of dark matter which can not be observed directly
although its gravitational effects on visible matter validate its
presence. In respect of recent WMAP data \cite{hinshaw}, our
universe is composed of 72 \% dark energy, 23 \% dark matter, and
5 \% ordinary (visible) matter.

Up to now, the most popular candidate of the dark energy is the
cosmological constant (vacuum energy) with the equation of state
parameter $\omega= -1$. However, the observed vacuum energy
density is at least $120 $ orders of magnitude smaller than
predicted by particle physics. This is the so-called cosmological
constant problem. In order to solve this problem, alternative
models based on a dynamical cosmological constant $\Lambda $, with
a negative equation of state have been constructed. These models
include a scalar field with a slowly varying energy density. In
quintessence models, the scalar field which is minimally coupled
to gravity with an equation of state $\omega>-1 $ acts as dark
energy and a potential energy dominating over kinetic energy leads
to the accelerating expansion \cite{Copeland}. If the scalar field
has a non-canonical kinetic energy then we have k-essence models
\cite{k-essence}. On the other hand, phantom energy models with a
negative kinetic energy assert an equation of state parameter
$\omega<-1 $ \cite{Copeland}. Besides, string-theory inspired
quintom models have also been analyzed. A model which includes the
combination of two-scalar fields have been considered \cite{feng},
one corresponding for the early time quintessence dominance,
$\omega>-1 $ and the other one corresponding for the late time
dominance, $\omega<-1 $. In addition, another string inspired
quintom model where tachyon is non-minimally coupled to gravity
obtained the conditions required for $\omega $ crosses over $ -1$
\cite{quintom}. Modified gravity models in the framework of
scalar-tensor theories have also been analyzed to explain the
acceleration of the universe \cite{fr}. A special case of these
type of models is the Brans-Dicke-Jordan-Thirry
\cite{jordan}-\cite{brans} theory where the curvature scalar
occurs only linearly in the lagrangian density. Whether the
quintessence field can be identified with the
Brans-Dicke-Jordan-Thirry field is an interesting question
\cite{sen2}-\cite{arik06}. In addition to explaining dark matter,
BD theory may have other advantages. In particular it has been
remarked that BD theory can be imbedded in electroweak theory
\cite{chernodub} and it can explain the cosmic coincidence problem
\cite{carneiro}. There exist a number of studies on accelerated
models in BD theory \cite{arias}- \cite{das}. For example, Sen
\textit{et al} \cite{sen} have found the potential relevant to
power law expansion in BD cosmology. In addition, in a work of
Setare \cite{setare}, the lower bound of $ \omega_{\Lambda}$ was
found $-0.9 $ using a holographic dark energy model in the
framework of non-flat BD cosmology.

In standard cosmology the rate of expansion of the universe
strongly depends on the equation of state of the matter-energy
that fills it. One immediate question which arises is that whether
there is any consistent modification of Einstein's equations such
that the expansion of the universe is independent of its content.
In the previous works of Arik, Calik and Sheftel
\cite{arik1}-\cite{arik06}, it is shown that BD scalar tensor
theory of gravity with the standard mass term potential
$(1/2)m^2\phi^2$ is capable of explaining the rapid primordial and
slow late-time inflation and a linearized non-vacuum late time
solution well accounts for the contribution of dark energy to the
Friedmann Equation, however, it does not account for the
contribution of dark matter. In this regard, we particularly focus
on the model consists of a modified Brans-Dicke-Jordan-Thirry
\cite{jordan}-\cite{brans} model where both the signs of the
kinetic term ($\phi \phi ^{\ast }=\phi _{1}^{2}-\phi _{2}^{2}$) in
its $\phi_{2}^{2}$ part and potential term bring a minus sign. The
models with this sign convention in Lagrangian have been termed as
quintom models \cite{feng}, \cite{quintom}, a word induced from quintessence
and phantom. We add an imaginary part to the BD field such as
$\phi=\phi_{1}+i\phi_{2}$, and search for a contribution to dark
matter in the presence of the imaginary part of $\phi$ field,
$\phi_{2}$.


\section{Field Equations}

In this work, we will show that both the dark matter contribution $\Omega _{%
\mathrm{DM}}$ and dark energy contribution $\Omega _{\Lambda }$ to
Friedmann Equation can be explained solely by BD theory of gravity
provided that BD scalar field is modified suitably. The most
straightforward modification is choosing the BD field as a complex
field defined by
\begin{equation}
\phi =\phi _{1}+i\phi _{2}=\phi _{\mathrm{R}}e^{i\beta }\label{comp ifi}
\end{equation} where $\phi _{\mathrm{R}}$ is real scalar field amplitude.
Such complex BD field can also be represented as in matrix form
\begin{equation}
\phi =\left(\begin{array}{cc}
\phi _{1} & \phi _{2} \\
-\phi _{2} & \phi _{1} \\
\end{array}\right)  =\phi _{1}+i\sigma _{2}\phi _{2}   \label{comp fi}
\end{equation}%
where $\sigma _{2}$ is a Pauli spin matrix.

However, we will take the phase of $\phi $ to be hyperbolic by replacing the term $%
i\beta =\Psi $ in (\ref{comp ifi}) such that $\phi $ becomes
\begin{equation}
\phi =\left(
\begin{array}{cc}
\phi _{1} & \phi _{2} \\
\phi _{2} & \phi _{1} \\
\end{array}%
\right)  \label{hyper fi}
\end{equation}%
and its conjugate matrix becomes%
\begin{equation}
\phi ^{\ast } =\left(
\begin{array}{cc}
\phi _{1} & -\phi _{2} \\
-\phi _{2} & \phi _{1} \\
\end{array}%
\right)  \label{hyper fi}
\end{equation}%
where
\begin{eqnarray}
\phi _{1} &=&\phi _{R}\cosh \Psi \label{fulya1}  \\
\phi _{2} &=&\phi _{R}\sinh \Psi \label{fulya2}
\end{eqnarray}%
where $\Psi$ is real. With this modification, we note here that
$\Psi $ gains a "Quintom" character since its kinetic contribution
($\phi \phi ^{\ast }=\phi _{1}^{2}-\phi _{2}^{2}$) brings a minus
sign. But nevertheless, in this paper, we will show that a
cosmological vacuum solution with flat space-like section is
capable of explaining how the Hubble parameter $H$ evolves with
the scale size of the universe $a(t)$ and how the solution of
fractional rate of change of BD scalar field, F contributes to the
evolution of $H$ in the late era in which the universe is
expanding at a slow rate. In the context of BD theory \cite{brans}
with self interacting potential and matter field, the action in
the
canonical form for real BD scalar field in our notation is given by%
\begin{equation}
S=\int d^{4}x\,\sqrt{g}\,\left[ -\frac{1}{8\omega }\,\phi _{1}^{2}\,R+\frac{1%
}{2}\,g^{\mu \upsilon }\,\partial _{\mu }\phi _{1}\,\partial _{\nu }\phi
_{1}-\frac{1}{2}m^{2}\phi _{1}^{2}+L_{M}\right] ,  \label{action*}
\end{equation}%
however, since we have modified the scalar BD field $\phi $ as in (\ref%
{hyper fi}), we also modify the action above as
\begin{equation}
S=\frac{1}{2}
\tr\int d^{4}x\sqrt{g}\left[ \frac{-1}{8\omega }\phi
\phi ^{\ast }R+\frac{1}{2}g^{\mu \nu }\partial _{\mu }\phi \partial _{\nu
}\phi ^{\ast }-\frac{1}{2}m^{2}\phi \phi ^{\ast }+IL_{M}\right] .
\label{actionmain}
\end{equation}%
where ${I}$ is the unit matrix. In particular we may expect that $\phi $ is
spatially uniform, but varies slowly with time. The signs of the non-minimal
coupling term and the kinetic energy term are properly adopted to $(+---)$
metric signature. In units where $c=\hbar =1$, we define Planck-length, $L_{%
\mathrm{p}}$, in such a way that $L_{\mathrm{P}}^{2}\phi _{\mathrm{R}%
}^{2}=\omega /2\pi $ where $\phi _{\mathrm{R}}$ is the present value in (\ref{fulya1},\ref{fulya2}). Hence the dimension of the scalar field is chosen to
be $L_{\mathrm{p}}^{-1}$, so that $G_{eff}$ has a dimension $L_{\mathrm{P}%
}^{2}$ since nonminimal coupling term $\,\phi _{R}^{2}\,R$ where $R$ is the
Ricci scalar, replaces with the Einstein-Hilbert term $\frac{1}{G_{N}}R$ in
such a way that $G_{eff}^{-1}=\frac{2\pi }{\omega }\phi _{R}^{2}$ where $%
G_{eff}$ is the effective gravitational constant as long as the dynamical
scalar field $\phi $ varies slowly with time. To be in accordance with the
weak equivalence principle, the matter part of the Lagrangian, $L_{M}$, is
decoupled from $\phi $ such that we have  considered the energy-momentum
tensor $T_{\nu }^{\mu }=diag\left( \rho ,-p,-p,-p\right) $ just with classical
perfect fluid where $\rho $ is the energy density, $p$ is the pressure. The
gravitational field equations derived from the variation of the action (\ref%
{actionmain}) with respect to Robertson- Walker metric is

\begin{equation}
\frac{3}{4\omega }\left( \frac{\dot{a}^{2}}{a^{2}}+\frac{k}{a^{2}}\right) -%
\frac{1}{2}\frac{\dot{\phi}\dot{\phi}^{\ast }}{\phi \phi ^{\ast }}+\frac{3}{%
4\omega }\left( \frac{\dot{a}}{a}\right) \left( \frac{\dot{\phi}\phi ^{\ast
}+\phi \dot{\phi}^{\ast }}{\phi \phi ^{\ast }}\right) -\frac{1}{2}m^{2}=
\frac{\rho_{M}}{\phi \phi ^{\ast }}  \label{ro}
\end{equation}%
\begin{eqnarray}
\lo -\frac{1}{4\omega }\left( \frac{2\ddot{a}}{a}+\frac{\dot{a}^{2}}{a^{2}}+%
\frac{k}{a^{2}}\right) -\left( \frac{1}{2}+\frac{1}{2\omega }\right) \frac{%
\dot{\phi}\dot{\phi}^{\ast }}{\phi \phi ^{\ast }}-\frac{1}{2\omega }\left(
\frac{\dot{a}}{a}\right) \left( \frac{\dot{\phi}\phi ^{\ast }+\phi \dot{\phi}%
^{\ast }}{\phi \phi ^{\ast }}\right)  \\ \label{press}
-\frac{1}{4\omega }\left( \frac{\ddot{\phi}\phi ^{\ast }+\ddot{\phi ^{\ast }}%
\phi }{\phi \phi ^{\ast }}\right) +\frac{1}{2}m^{2}=\frac{p_{M}}{\phi \phi ^{\ast }} \nonumber
\end{eqnarray}%
\begin{equation}
\frac{\ddot{\phi}}{\phi }+3\left( \frac{\dot{a}}{a}\right) \frac{\dot{\phi}}{%
\phi }+m^{2}-\frac{3}{2\omega }\left( \frac{\ddot{a}}{a}+\frac{\dot{a}^{2}}{%
a^{2}}+\frac{k}{a^{2}}\right) =0  \label{fieqn}
\end{equation}%
where $k$\ is the curvature parameter with $k=-1$, $0$, $1$\ corresponding
to open, flat, closed universes respectively and $a\left( t\right) $ is the
scale factor of the universe (dot denotes $\frac{d}{dt}$). Since in the
standard theory of gravitation, the total energy density $\rho $ is assumed
to be composed of $\rho =\rho _{\Lambda }+\rho _{M}$ where $\rho _{\Lambda }$
is the energy density of the universe due to the cosmological constant which
in modern terminology is called as \textquotedblleft \textit{dark
energy\textquotedblright ,} the right hand sides of (\ref{ro}, \ref{press})
are adopted to the matter energy density term $\rho _{M}$ instead of $\rho $
and $p_{M}$ instead of $p$ where $M$ denotes everything except the $\phi $
field. The main reason behind doing such an organization is that whether if
the $\phi $ terms on the left-hand side of (\ref{ro}) can accommodate a
contribution to due to what is called dark matter. In addition, the right
hand side of the $\phi $ equation (\ref{fieqn}) is set to be zero according
to the assumption imposed on the matter Lagrangian $L_{M}$ being independent
of the scalar field $\phi $. For the vacuum ($\rho =p=0$) and flat space
like ($k=0$) section solutions, we start with defining the fractional rate of change of $%
\phi $ as
\begin{equation}
F\left( a\right) =\frac{\dot{\phi}\phi ^{\ast }}{\phi \phi ^{\ast }}=\left(
\begin{array}{cc}
F_{1} & F_{2} \\
F_{2} & F_{1} \\

\end{array}%
\right)   \label{F}
\end{equation}
where
\begin{equation}
F_{1} =\frac{\dot{\phi _{1}}\phi _{1}-\dot{\phi _{2}}\phi _{2}}{\phi
_{1}^{2}-\phi _{2}^{2}}=\frac{\dot{\phi}_{R}}{\phi _{R}}, \\ \bigskip
F_{2} =\frac{\dot{\phi _{2}}\phi _{1}-\dot{\phi _{1}}\phi _{2}}{\phi
_{1}^{2}-\phi _{2}^{2}}=\dot{\Psi}  \label{compf1andf2}
\end{equation}
and the Hubble parameter as $H\left( a\right) =\dot{a}/a$, hence, we
rewrite the left hand-side of the field equations (\ref{ro}-\ref{fieqn})
in terms of $H(a)$, $F_{1}(a)$, $F_{2}(a)$ and their
derivatives with respect to the scale size of an universe $a$, as
\begin{equation}
3H^{2}\mathbf{-}2\omega F_{1}^{2}\mathbf{+}2\omega F_{2}^{2}\mathbf{+}6F_{1}H%
\mathbf{-}2\omega m^{2}=0  \label{R}
\end{equation}

\begin{equation}
3H^{2}+\left( 2\omega +4\right) F_{1}^{2}-2\omega
F_{2}^{2}+4F_{1}H+2aHF_{1}^{\prime }+2aHH^{\prime }-2\omega m^{2}=0
\label{P}
\end{equation}

\begin{equation}
-6H^{2}+2\omega F_{1}^{2}\mathbf{+}2\omega F_{2}^{2}+6\omega F_{1}H\mathbf{+}%
2\omega aHF_{1}^{\prime }-3aHH^{\prime }+2\omega m^{2}=0  \label{FReel}
\end{equation}

\begin{equation}
\left( 4\omega F_{1}\mathbf{+}6\omega H\right) F_{2}+2\omega aHF_{2}^{\prime
}=0  \label{Fim}
\end{equation}
where prime denotes $\frac{d}{da}$. Since solving these coupled equations
analytically is hard enough, we have put forward \ following perturbation
solution as%
\begin{equation}
H=H_{\infty }+H_{1}\left( \frac{a_{0}}{a}\right) ^{\alpha }+H_{2}\left(
\frac{a_{0}}{a}\right) ^{2\alpha }  \label{per1}
\end{equation}%
\begin{equation}
F_{1}=F_{1\infty }+F_{11}\left( \frac{a_{0}}{a}\right) ^{\alpha
}+F_{12}\left( \frac{a_{0}}{a}\right) ^{2\alpha }  \label{per2}
\end{equation}%
\begin{equation}
F_{2}=F_{2\infty }+F_{21}\left( \frac{a_{0}}{a}\right) ^{\alpha
}+F_{22}\left( \frac{a_{0}}{a}\right) ^{2\alpha }  \label{per3}
\end{equation}%
where $H_{\infty }$, $H_{1}$, $H_{2}$, $F_{1\infty }$, $F_{11}$, $F_{12}$, $%
F_{2\infty }$, $F_{21}$, $F_{22}$ are perturbation constants and $\alpha $
is an exponential factor to be determined. With the transformation
\begin{equation}
u=\left( \frac{a_{0}}{a}\right) ^{\alpha },  \label{utrans}
\end{equation}%
(\ref{R}-\ref{Fim}) becomes
\begin{equation}
3H^{2}-2\omega F_{1}^{2}+2\omega F_{2}^{2}+6HF_{1}-2\omega m^{2}=0
\label{rovaccum}
\end{equation}%
\begin{equation}
\fl3H^{2}+\left( 2\omega +4\right) F_{1}^{2}-2\omega
F_{2}^{2}+4HF_{1}-2\alpha uH\left( \frac{dF_{1}}{du}\right) -2\alpha
uH\left( \frac{dH}{du}\right) -2\omega m^{2}=0  \label{pvaccuum}
\end{equation}

\begin{equation}
\fl-6H^{2}+2\omega F_{1}^{2}+2\omega F_{2}^{2}+6\omega HF_{1}-2\omega \alpha
uH\left( \frac{dF_{1}}{du}\right) +3\alpha uH\left( \frac{dH}{du}\right)
+2\omega m^{2}=0  \label{fivaccum}
\end{equation}%
\begin{equation}
-2\omega \alpha uH\left( \frac{dF_{2}}{du}\right) +\left( 4\omega F_{1}%
\mathbf{+}6\omega H\right) F_{2}=0  \label{fivaccum*}
\end{equation}%
and (\ref{per1}-\ref{per3}) becomes%
\begin{equation}
H=H_{\infty }+H_{1}u+H_{2}u^{2}  \label{p1}
\end{equation}%
\begin{equation}
F_{1}=F_{1\infty }+F_{11}u+F_{12}u^{2}  \label{p2}
\end{equation}%
\begin{equation}
F_{2}=F_{2\infty }+F_{21}u+F_{22}u^{2}.  \label{p3}
\end{equation}

Substituting (\ref{p1}-\ref{p3}) into (\ref{rovaccum}-\ref{fivaccum*}) and
keeping only the zeroth, first and second order terms of u and neglecting
higher order terms of $u$, we get the following equations to be solved. In the
zeroth order of $u$;
\begin{equation}
3H_{\infty }^{2}\mathbf{-}2\omega F_{1\infty }^{2}\mathbf{+}2\omega
F_{2\infty }^{2}\mathbf{+}6F_{1\infty }H_{\infty }\mathbf{-}2\omega m^{2}=0
\label{fivaccuma}
\end{equation}%
\begin{equation}
3H_{\infty }^{2}+\left( 2\omega +4\right) F_{1\infty }^{2}-2\omega
F_{2\infty }^{2}+4F_{1\infty }H_{\infty }-2\omega m^{2}=0  \label{q2}
\end{equation}%
\begin{equation}
-6H_{\infty }^{2}+2\omega F_{1\infty }^{2}\mathbf{+}2\omega F_{2\infty
}^{2}+6\omega F_{1\infty }H_{\infty }\mathbf{+}2\omega m^{2}=0  \label{q3}
\end{equation}%
\begin{equation}
\lbrack 4\omega F_{1\infty }\mathbf{+}6\omega H_{\infty }]F_{2\infty }=0
\label{q4}
\end{equation}%
in the first order of $u$;
\begin{equation}
(6F_{1\infty }+6H_{\infty })H_{1}+(6H_{\infty }\mathbf{-}4\omega F_{1\infty
})F_{11}+4\omega F_{21}F_{2\infty }=0  \label{q5}
\end{equation}%
\begin{equation}
\fl((6\mathbf{-}2\alpha )H_{\infty }+4F_{1\infty })H_{1}+((4\mathbf{-}2\alpha
)H_{\infty }+\left( 4\omega +8\right) F_{1\infty })F_{11}-4\omega
F_{21}F_{2\infty }=0  \label{q6}
\end{equation}%
\begin{equation}
\fl((3\alpha -12)H_{\infty }+6\omega F_{1\infty })H_{1}+((6\omega \mathbf{-}%
2\omega \alpha )H_{\infty }+4\omega F_{1\infty })F_{11}+4\omega
F_{21}F_{2\infty }=0  \label{aq}
\end{equation}%
\begin{equation}
\lbrack (\mathbf{-}2\omega \alpha +6\omega )H_{\infty }+4\omega F_{1\infty
}]F_{21}+4\omega F_{11}F_{2\infty }+6\omega H_{1}F_{2\infty }=0  \label{q}
\end{equation}%
in the second order of $u$;
\begin{eqnarray}
\fl3H_{1}^{2}+6H_{\infty }H_{2}-2\omega F_{11}^{2}-4\omega F_{1\infty }F_{12}
\label{s1} \\
+4\omega F_{2\infty }F_{22}+2\omega F_{21}^{2}+6F_{1\infty
}H_{2}+6F_{11}H_{1}+6F_{12}H_{\infty } =0  \nonumber
\end{eqnarray}%
\begin{eqnarray}
\fl(3-2\alpha )H_{1}^{2}\mathbf{\mathbf{+}}\left( 2\mathbf{\omega +}4\right)
F_{11}^{2}+(4-2\alpha )F_{11}H_{1}+(4F_{1\infty }+6H_{\infty }-4\alpha
H_{\infty })H_{2}  \label{s2} \\
+[4H_{\infty }+\left( 4\mathbf{\omega +}8\right) F_{1\infty }-4\alpha
H_{\infty }]F_{12}-2\omega F_{21}^{2}-4\omega F_{2\infty }F_{22} =0
\nonumber
\end{eqnarray}

\begin{eqnarray}
\fl(3\alpha -6)H_{1}^{2}\mathbf{\mathbf{+}}2\omega F_{11}^{2}+(-2\omega
\alpha +6\omega )F_{11}H_{1}+(-12H_{\infty }+6\alpha H_{\infty }+6\omega
F_{1\infty })H_{2}  \label{s3} \\
+(-4\omega \alpha H_{\infty }+4\omega F_{1\infty }+6\omega H_{\infty
})F_{12}+2\omega F_{21}^{2}+4\omega F_{2\infty }F_{22} =0  \nonumber
\end{eqnarray}

\begin{eqnarray}
\fl(4\omega F_{1\infty }-4\omega \alpha H_{\infty }+6\omega H_{\infty
})F_{22}+(-2\omega \alpha H_{1}+4\omega F_{11}+6\omega H_{1})F_{21}
\label{s4} \\
+4\omega F_{2\infty }F_{12}+6\omega F_{2\infty }H_{2} =0.  \nonumber
\end{eqnarray}

\section{Solutions}

Solving the equation set (\ref{fivaccuma}-\ref{q4}) and (\ref{q5}-\ref{q})
 provide respectively,
\begin{equation}
F_{2\infty }=0\qquad H_{\infty }=\frac{2\left( \omega +1\right) \sqrt{\omega
}m}{\sqrt{\left( 6\omega ^{2}+17\omega +12\right) }}\qquad F_{1\infty }=%
\frac{H_{\infty }}{2\omega +2}  \label{Hinfva}
\end{equation}%
\begin{equation}
\alpha =3+\frac{1}{\omega +1}\qquad F_{11}=-\frac{3}{2}H_{1}\qquad F_{21}=%
\mathrm{free-parameter}  \label{alfaa}
\end{equation}%
and afterwards, substituting (\ref{Hinfva}, \ref{alfaa}) into the equation set
(\ref{s1}-\ref{s4}) yields the following equation set to be solved for $H_{2}
$, $F_{12}$, $F_{21}$, $F_{22}$ as
\begin{equation}
\fl(12\omega +18)H_{\infty }H_{2}+(8\omega +12)H_{\infty }F_{12}+4\omega
(\omega +1)F_{21}^{2}=(9\omega ^{2}+21\omega +12)H_{1}^{2}  \label{c1}
\end{equation}%
\begin{equation}
\fl-(12\omega 16)H_{\infty }H_{2}-(12\omega +16)H_{\infty }F_{12}-4\omega
(\omega +1)F_{21}^{2}=-(9\omega ^{2}+27\omega +20)H_{1}^{2}  \label{c2}
\end{equation}%
\begin{equation}
\fl(18\omega +24)H_{\infty }H_{2}-(12\omega ^{2}+16\omega )H_{\infty
}F_{12}+4\omega (\omega +1)F_{21}^{2}=-(9\omega ^{2}+21\omega +12)H_{1}^{2}
\label{c3}
\end{equation}%
\begin{equation}
F_{22}=-\frac{H_{1}}{H_{\infty }}F_{21}.  \label{c4}
\end{equation}

\bigskip
To proceed one step further, we write the standard Friedmann
equation:
\begin{equation}
\left( \frac{H}{H_{0}}\right) ^{2}=\Omega _{\Lambda }+\Omega _{M}\,\left(
\frac{a_{0}}{a}\right) ^{3}  \label{frd in Brans-dicke}
\end{equation}%
and we fit all theory parameters to the observational density parameters;
\begin{equation}
\Omega _{\Lambda }=\frac{H_{\infty }^{2}}{H_{\Sigma }^{2}},
\label{omegalamda}
\end{equation}%
\begin{equation}
\Omega _{M}=\frac{2H_{\infty }H_{1}}{H_{\Sigma }^{2}},  \label{omegamatter}
\end{equation}%
where%
\begin{equation}
H_{\Sigma }^{2}=H_{\infty }^{2}+2H_{\infty }(H_{1}+H_{2})+H_{1}^{2}.
\label{omegasigma}
\end{equation}%
With these relations above and the constraint $\Omega _{\Lambda }$ $+$ $%
\Omega _{M}=1$, where $\Omega _{M}=\Omega _{\mathrm{VM}}+\Omega _{\mathrm{DM}%
}$, we can express theoretical parameters $H_{1}$ in terms of the
observational density parameters $\Omega _{\Lambda }$, $\Omega _{M}$ and $%
H_{\infty }$ as%
\begin{equation}
H_{1}=\frac{\Omega _{M}}{2\Omega _{\Lambda }}H_{\infty }  \label{he*}
\end{equation}%
Using recent observational results \cite{hinshaw} on density parameters $%
\Omega _{DM}\simeq 0.28$, $\Omega _{\Lambda }\simeq 0.72$ and $\Omega _{%
\mathrm{VM}}=0$ (since the universe we study in this theory is vacuum)
together with (\ref{he*}) we determine;
\begin{equation}
H_{1}=\frac{0.28}{1.44}H_{\infty }\simeq 0.19H_{\infty }.  \label{H1 delta=0}
\end{equation}%
Similarly, when we solve the equations (\ref{c1}-\ref{c4});\ the solutions
are;

\begin{equation}
\fl H_{2}H_{\infty }=\frac{1}{34\omega +12\omega ^{2}+24}\left( 18\omega
^{2}H_{1}^{2}-8\omega ^{2}F_{21}^{2}-4\omega
^{3}F_{21}^{2}-12H_{1}^{2}-\omega H_{1}^{2}-4\omega F_{21}^{2}+9\omega
^{3}H_{1}^{2}\right)\label{a}
\end{equation}%
\begin{equation}
\fl F_{12}H_{\infty }=\frac{1}{68\omega +24\omega ^{2}+48}\left(
84H_{1}^{2}-4\omega ^{2}F_{21}^{2}-4\omega F_{21}^{2}+123\omega
H_{1}^{2}+45\omega ^{2}H_{1}^{2}\right).\label{b}
\end{equation}

As $\omega \rightarrow \infty $;%
\begin{equation}
H_{2}\simeq \left( -\frac{1}{3}\omega \frac{F_{21}^{2}}{H_{\infty }}+\frac{3%
}{4}\omega \frac{H_{1}^{2}}{H_{\infty }}\right)\label{c}
\end{equation}%
\begin{equation}
F_{12}\simeq \left( -\frac{4}{24H_{\infty }}F_{21}^{2}+\frac{45}{24H_{\infty }}%
H_{1}^{2}\right)\label{d}
\end{equation}%
\begin{equation}
F_{22}=-\frac{H_{1}}{H_{\infty }}F_{21}\label{e}
\end{equation}

At this point, we emphasize that $H_{2}$ must be zero in order to make sense
with recent observational data on density parameters of the universe and to
find the exact value for $F_{21}$. Therefore, we insert $H_{2}=0$ and we get
\begin{equation}
F_{21}\simeq 0.28H_{\infty }\label{fd}
\end{equation}

\begin{equation}
F_{12}\simeq0.05H_{\infty }\label{f}
\end{equation}%
\begin{equation}
F_{22}\simeq-0.06H_{\infty }.\label{g}
\end{equation}

Hence, with these perturbation constants (\ref{Hinfva}, \ref{alfaa}, \ref{H1 delta=0}, \ref{fd}-\ref{g}) found from theory we can express (\ref{per1}-\ref{per3});
\begin{equation}
H=H_{\infty }+0.19H_{\infty }\left( \frac{a_{0}}{a}\right) ^{3}  \label{M1}
\end{equation}%
\begin{equation}
F_{1}=\frac{H_{\infty }}{2\omega +2}-0.28H_{\infty }\left( \frac{a_{0}}{a}%
\right) ^{3}+0.05H_{\infty }\left( \frac{a_{0}}{a}\right) ^{6}  \label{M2}
\end{equation}%
\begin{equation}
F_{2}=0.28H_{\infty }\left( \frac{a_{0}}{a}\right) ^{3}-0.06H_{\infty }\left(\frac{%
a_{0}}{a}\right)^{6}  \label{M3}
\end{equation}%
where
\begin{equation}
H_{\infty }\simeq 0.84H_{0}  \label{Hinf delta=0}
\end{equation}%
if (\ref{M1}) is satisfied for $H=H_{0}$, and $H_{0}$ is the present value
of the Hubble parameter \cite{hinshaw}.

\bigskip

\section{Conclusion}

In this paper, we have analyzed the dark matter $\Omega _{DM}$ and
dark energy contribution $\Omega _{\Lambda }$ to Friedmann
Equation solely by modified BD theory of gravitation with no other
input. As far as we know, the scalar field $\phi $ was always
examined individually, however, we brought forward a new idea such
that it can have different components and each of these components
can account for different energy densities.

Actually, the starting
point of our motivation originates from this point in the sense
that when BD theory of gravitation with solely scalar field $\phi$
is substituted into role as discussed in our previous work \cite{arik06}, we
have shown that WMAP+SnIa data \cite{hinshaw}, \cite{riess}-\cite{WMAP2} favor
this model instead of the standard Einstein cosmological model
with cosmological constant (LCDM model \cite{WMAP}-\cite{WMAP2}) under the condition that
the new density parameter $\Omega _{_{\Delta }}$ induced in
Friedmann equation in standard cosmology to be $\Omega _{_{\Delta
}}<0$ and $H_{2}$ seen in the equation (\ref{p1}) to be $H_{2}<0$
instead of $H_{2}=0$. (for further information, see
\cite{arik06}). In other words, at this stage, we have realized
that the more we force $H_{2}$ to be less than zero in the real
phase of the model with individual scalar field $\phi$, it fits
WMAP+SnIa data much more confidentially than LCDM model \cite{WMAP}-\cite{WMAP2}. To do
this, in the first attempt, we have used a complex scalar field
$\phi =\phi _{1}+i\phi _{2}$ so that WMAP+SnIa \cite{hinshaw}, \cite{riess}-\cite{WMAP2} data will favor the
model with modified BD field $\phi $ in its complex phase.
Although this approach has brought brand new considerations and
aspects to Friedmann Equation in the concept of dark matter and
dark energy, a more suitable solution was found with the
modification of scalar field by using a hyperbolic phase $i\beta
=\Psi $. Having solved the field equations including the
hyperbolic phase, we achieved the field equations
(\ref{R}-\ref{Fim}) of modified BD scalar tensor theory namely equations of ''Quintom'' model.

 To solve these above field equations, we
put up the argument of perturbative solutions with the constant terms $%
H_{\infty },H_{1},H_{2},F_{1\infty ,}F_{11},F_{12},F_{2\infty },F_{21}$ and $%
F_{22}.$ All of these constants have made it possible to originate new
predictions on dark matter and dark energy contribution of BD\ theory.

To begin with, the most significant evidence for the idea that this
modification needs real attention is the solution of $\alpha .$ It can easily
be seen that, when $\omega \rightarrow \infty $ (where BD approaches
Einstein theory), $\alpha \rightarrow 3,$ as it appears in the Friedman
Equation in the form $\Omega _{M}\left( \frac{a_{0}}{a}\right) ^{3}$.
Similarly, the term $H_{\infty }$ which has no scale factor term, just like
the energy density term due to the cosmological constant $\Omega _{\Lambda }$
in the Friedman Equation, was found purely from theory;

\begin{equation}
H_{\infty }=\frac{2\left( \omega +1\right) \sqrt{\omega
}m}{\sqrt{\left( 6\omega ^{2}+17\omega +12\right) }}.  \label{Hinfv}
\end{equation}%
From the equation (\ref{M1}) in the equation set (\ref{M1}-\ref{M3}),
we see that the second term is found to be smaller than the first one
and the third term is found to be smaller than the second one. Namely, the dominating
term is the first one which can be interpreted as the contribution to dark
energy. On the other hand, the second term can be considered as the contribution to dark matter.
However, the situation is different for $F_{1}$ and $F_{2}$ as it is seen in the equations
 (\ref{M2}, \ref{M3}). As it was mentioned before, in the absence of $F_{2}$ term,
 where $F=F_{1}$, the theory was able to explain the contribution to dark energy but not to dark matter.
 Our aim was to find a contribution to dark matter in the presence of $F_{2}$, with the component $F_{21}$ since it is coupled
 with $(\frac{a_{0}}{a})^{3} $. Namely; it is
agreeable to predict that while $F_{1\infty }$ which is not
coupled with a scale factor term is contributing to dark energy,
$F_{21}$ is contributing to dark matter. Hence, the introduction
of a hyperbolic phase for BD field results in a flat vacuum
cosmological solution of Hubble parameter H and fractional rate of
change
 of BD scalar field, F which asymptotically approach constant values. At late stages, hyperbolic phase of BD field behaves
like dark matter.
\section{Acknowledgments}
We would like to thank the anonymous referee for thoughtful comments and
valuable suggestions on this paper. Besides,
    we would like to thank our immortal teacher, Prof. Engin Arik, for her valuable suggestions and contributions to this paper. This work is partially supported by Bogazici University Research Fund
    and by Turkish Atomic Energy Authority (TAEK).

\end{document}